\documentclass[12pt]{article}
\let\section=\subsection     \let\subsection=\subsubsection                
\usepackage{graphicx}

\begin{document}
\begin{center}
   {\large \bf HADRON-HADRON INTERACTIONS IN THE CONSTITUENT
   QUARK MODEL: RESULTS AND EXTENSIONS}\\[5mm]
   E.S. Swanson\\[5mm]
   {\small \it  Department of Physics and Astronomy \\
   University of Pittsburgh, Pittsburgh PA 15260\\
   and Jefferson Lab, 12000 Jefferson Ave, Newport News VA 23606 }
\end{center}

\begin{abstract}\noindent
Hadronic interactions are discussed within the context of the constituent
quark model. The ``Quark Born Diagram" methodology is outlined, extensive
applications to meson-meson and meson-baryon interactions are discussed, and 
general features of these interactions are highlighted. The second half of
this document deals with shortcomings of the quark model approach and
methods to overcome them. These include relativistic kinematics, unitarity,
nonlocal potentials, coupled channel
effects, and the chiral nature of the pion.
\end{abstract}

\section{Ubiquitous Hadronic Interactions}

A microscopic understanding hadron-hadron scattering remains an elusive goal of hadronic physics. 
This is unfortunate because the interactions of hadrons is important from a variety of
perspectives. At the hadronic level, it provides  vital  insight into the dynamics of quarks
and gluons.  It is also relevant to the search for the quark gluon plasma, since 
hadronic interactions can mask putative signals for the QCD phase transition. Applications
extend beyond hadronic physics: for example the search for CP violating phases in the
final states of $D$ of $B$ decays will require correctly accounting for strong interaction final state phases\cite{foot}. Of course, hadron-hadron interactions are directly relevant to a longstanding goal of
nuclear physics -- deriving the nuclear force from QCD.  This issue is not just of intellectual
concern since it is important to be able to extend our understanding of internuclear forces
to extreme conditions (of temperature and pressure) so that a variety of astrophysical and 
cosmological issues may be reliably examined.

Given the importance of the area, it is not surprising that a number of techniques have been
developed to address the dynamics of hadron-hadron interactions. These roughly fall into two classes:
those which treat hadrons as elementary fields and those which attempt to describe the interactions
using QCD as the starting point. Among the former are potential approaches\cite{potl} (Bonn, Paris, Argonne),
relativistic hydrodynamic approaches\cite{rel}, and a variety of 
effective field theories\cite{eff} (chiral perturbation theory, effective NN theory). 
Among the latter are  lattice gauge theory\cite{lattice},
Schwinger-Dyson models\cite{SD}, light cone field theory\cite{Lepage:1980fj}, a multigluon 
dipole interaction model\cite{peskin}, and constituent quark models.

Unfortunately, it is not possible to summarize the advantages and faults of all of these methods
here.  Rather, the remainder of this document focuses on attempts to describe hadronic 
interactions which are based on the nonrelativistic constituent quark model (CQM).  The CQM
is beyond a doubt the most widely used description of the static properties of hadrons -- 
largely because it is
able to describe hundreds of experimental data with a handful of parameters
in a comparatively simple picture. Thus it is no surprise that attempts to describe 
hadronic interactions with the CQM have a long history \cite{liberman}.  Despite the well-known
short comings of the CQM (some of which will be discussed below) the benefits are immediately
apparent: the CQM may be applied to {\it any} hadronic interactions 
and the predictions {\it are essentially parameter-free}. By
this I mean that the parameters of the CQM are very strongly fixed by comparison with static
hadron properties -- there is no wiggle room when computing interactions! Of course the extension
of the CQM to dynamic properties of hadrons requires some extrapolation\footnote{For example, 
the dynamics of flux tubes may be essentially ignored in conventional meson and baryons. This is no longer
the case for multiquark systems.}; however, this should not
form a barrier in itself. To paraphrase Feynman, ``Trust your model and see how far it takes you."

\section{Quark Born Diagrams}

In the following we shall consider a CQM which includes Coulomb and linear central potentials,
a spin-spin colour hyperfine interaction, and possibly spin-orbit and tensor interactions. 
It may be somewhat of a surprise that the Breit-Fermi interactions are included here since
they are normally only required to 
achieve detailed agreement with spectroscopic 
and other static properties of the hadrons.  However, as we shall see shortly, 
subleading (in $v/c$) interactions can dominate hadronic scattering!

A great deal of effort has been expended on variational\cite{Weinstein:1990gu}
and resonating group 
approaches\cite{rgm} to scattering in the CQM. Here I describe a simplified approach
where one evaluates the T-matrix at Born order\cite{Barnes:1992em,Swanson:1992ec}, called the
``Quark Born Diagram" (QBD) method.
The T-matrix for meson-meson scattering is shown diagrammatically in Fig 1. The diagrams
represent momentum flow in the Born order term of the Neumann series and must be
attached to external mesonic wavefunctions. Quark exchange must occur to maintain 
asymptotic colour singlet states.
In doing  these computations one needs
to be aware that scattering with composite objects  can be very different than pointlike objects.
For example, hermiticity of the scattering amplitude is no longer trivial but requires that
the wavefunctions be exact eigenstates of $H_0$. This property is related to the ``post-prior"
discrepancy\cite{Swanson:1992ec,Schiff}. Furthermore, consistency of the kinematical relationships 
must be maintained with the Hamiltonian (nonrelativistic Hamiltonians require 
nonrelativistic kinematics to maintain hermiticity). Also, one cannot change parameters at will!
For example, mesonic radii or quark masses cannot be independently modified. One must instead 
adjust a Hamiltonian parameter, solve for the spectrum again, and then evaluate the T-matrix.

\begin{center}
   \includegraphics[width=8truecm]{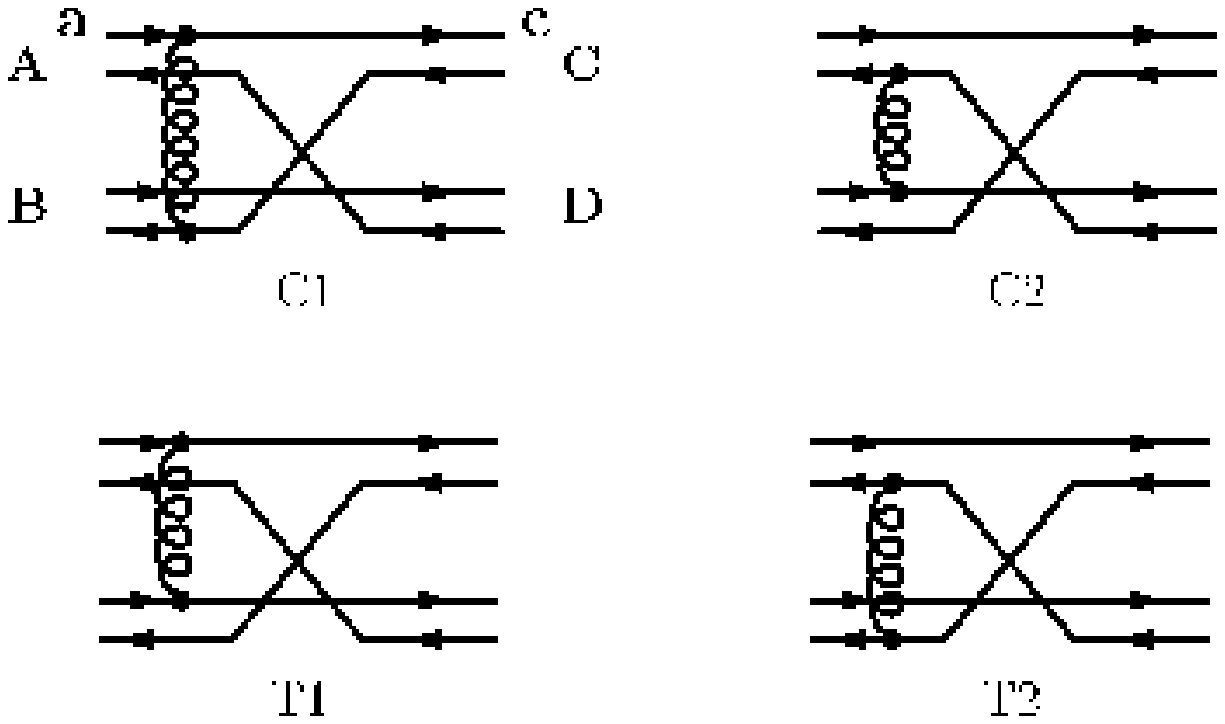}\\
   \parbox{14cm}
	{\centerline{\footnotesize 
	Fig.~1: Diagrams Contributing to Meson-Meson Scattering}}
\end{center}

\subsection{Applications}
The application of the Quark Born formalism requires some care. For example, if one is to 
restrict attention to the terms shown in Fig 1, then channels where strong quark annihilation
effects are expected (such as $I=0$ $\pi\pi$) should be avoided.  Furthermore, if the 
predicted  interactions are strong, the Born order results should be iterated (this is
discussed below). 
The Quark Born Diagram method has been applied to a variety of hadronic reactions. These include
I=2 $\pi\pi$ scattering\cite{Barnes:1992em,Barnes:2001hu} (with surprising agreement considering
the relativistic and chiral nature of the pion), I=3/2 $\pi K$ scattering\cite{Barnes:1992qa} (testing
Bose symmetry breaking due to the strange quark mass), $KN$ scattering\cite{Barnes:1994ca} (demonstrating surprising agreement in the S-wave and a dramatic failure in the P-wave\cite{foot2}), short range $NN$ scattering\cite{Barnes:1993nu} (in agreement with resonating group computations), $BB$ scattering
\cite{Barnes:1999hs} (in agreement with lattice computations), $J/\psi-\pi$ 
scattering\cite{Wong:2000zb} (in strong disagreement with previous estimates and in agreement with rudimentary data), $\pi\rho$ scattering\cite{Barnes:2001hu} (an examination of the generation of hadronic  spin orbit and tensor 
interactions from the quark level), and possible meson-meson bounds states\cite{Dooley:1992bg} (the 
$f_0(1710)$ may be identified as a $K^*\bar K^*- \omega\phi$ bound state).

A comparison of the predicted and experimental isotensor S-wave  $\pi\pi$ phase  shifts is shown 
in Fig 2.
A sample cross section prediction relevant to charm suppression at RHIC is shown in Fig 3.

\begin{center}
   \includegraphics[width=5.5cm,angle=-90]{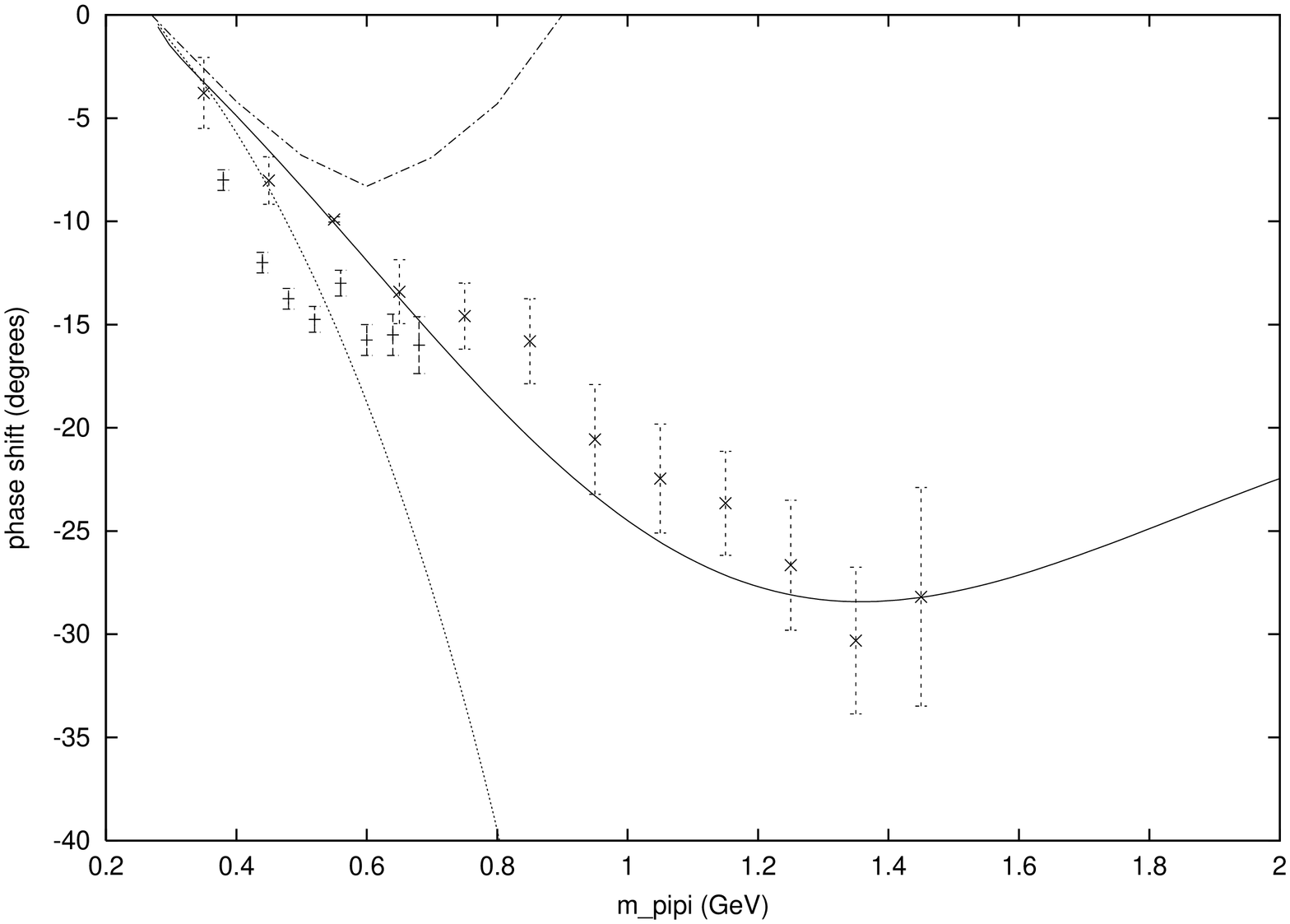}\\
   \parbox{14cm}
	{\centerline{\footnotesize 
	Fig.~2: The $I=2$ S-wave $\pi\pi$ Phase Shift. The solid line is the QBD 
        prediction\cite{Barnes:1992em},}}
	{\centerline{\footnotesize 
        the dashed line is the one loop chiral prediction\cite{Donoghue:1988xa}, 
        while the lower dashed line is}}
	{\centerline{\footnotesize 
        the tree order chiral result.
        Data are from \cite{pipidata}.}}
\end{center}

\begin{center}
   \includegraphics[width=5.5cm,angle=-90]{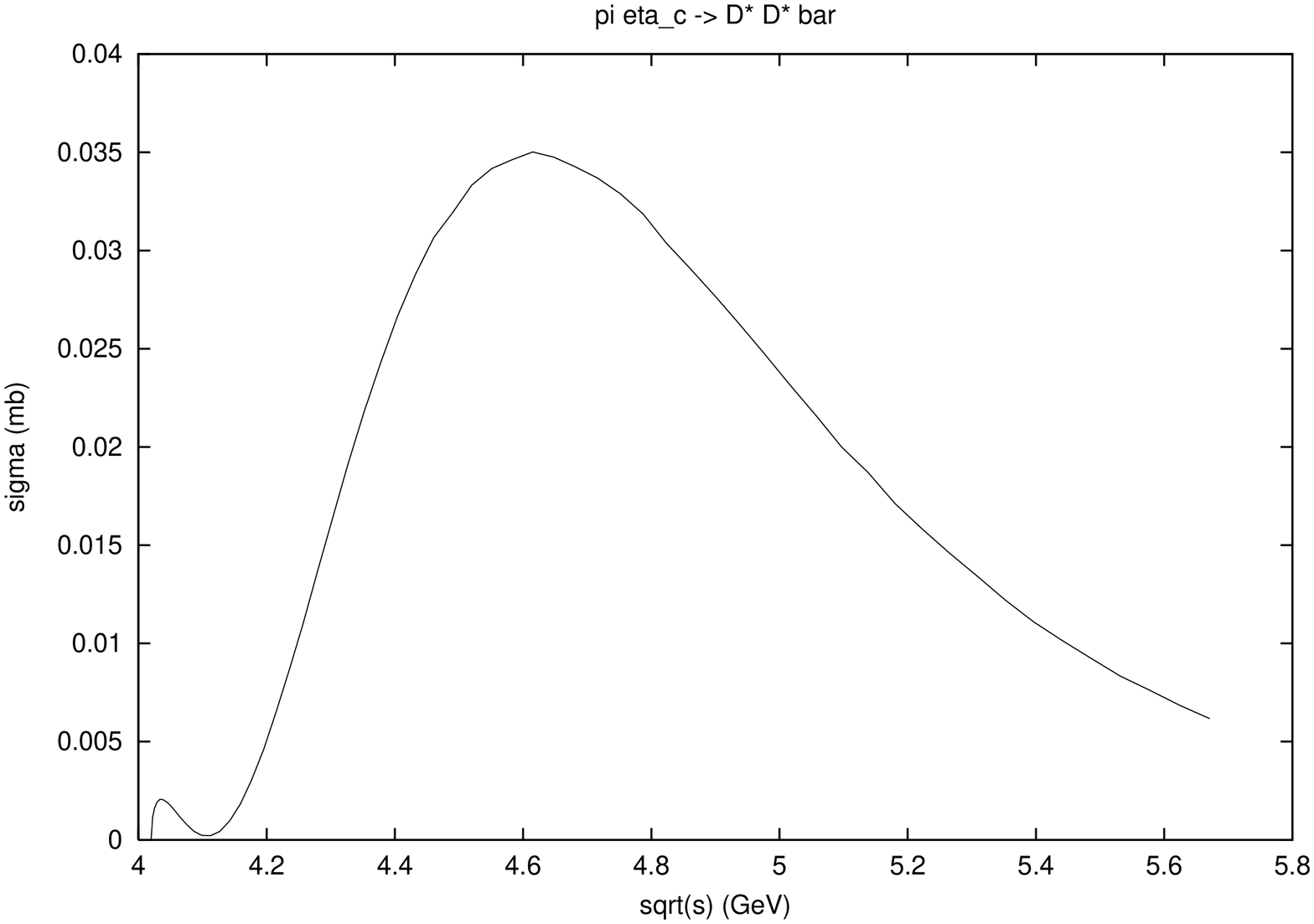}\\
   \parbox{14cm}
	{\centerline{\footnotesize 
	Fig.~3: Prediction of the $\pi \eta_c \rightarrow D^* \bar D^*$ Cross Section.}}
\end{center}

\section{Limitations and Extensions of the CQM}

As mentioned above, the CQM in general, and the QBD formalism in particular, suffers from
several inadequacies. We summarize several of these and discuss methods for addressing
them. 

\subsection{Unitarity, Relativistic Kinematics, and Nonlocality}

Perhaps the simplest problem arises when the scattering is so strong that the tree level
diagrams of the QBD method are inaccurate. This may be tested by comparing QBD predictions
to more complete resonating group calculations (which iterate the scattering to all orders
and can, in principle, include the effects of coupled channels, wavefunction distortion, etc).
Such a comparison of $\rho\rho$ scattering in isospin 2 was made in Ref. \cite{Swanson:1992ec} 
where the accuracy of the QBD results were explicitly demonstrated.  Nevertheless, it is worth
observing that strong interactions may be accounted for by extracting effective potentials for
the process in question and iterating the potential in the appropriate Schr\"odinger equation.
An example of this is shown in Fig 4, where the effective $\pi\pi$ interaction has been
extracted from the QBD $\pi\pi$ T-matrix and iterated in the nonrelativistic Schr\"odinger 
equation. This is shown as a dashed line. Evidently, the agreement with data is ruined by this
procedure (I am presenting the worst possible case -- the procedure
works very well in general). This is because the light mass of the pions (and the large
invariant masses at which the formalism is being applied) requires relativistic kinematics
to be employed (this has been used in the 
QBD prediction of Fig 2). 
Thus the nonrelativistic Schr\"odinger equation is inappropriate for iterating the 
effective $\pi\pi$ interaction.

\begin{center}
   \includegraphics[width=4cm,angle=-90]{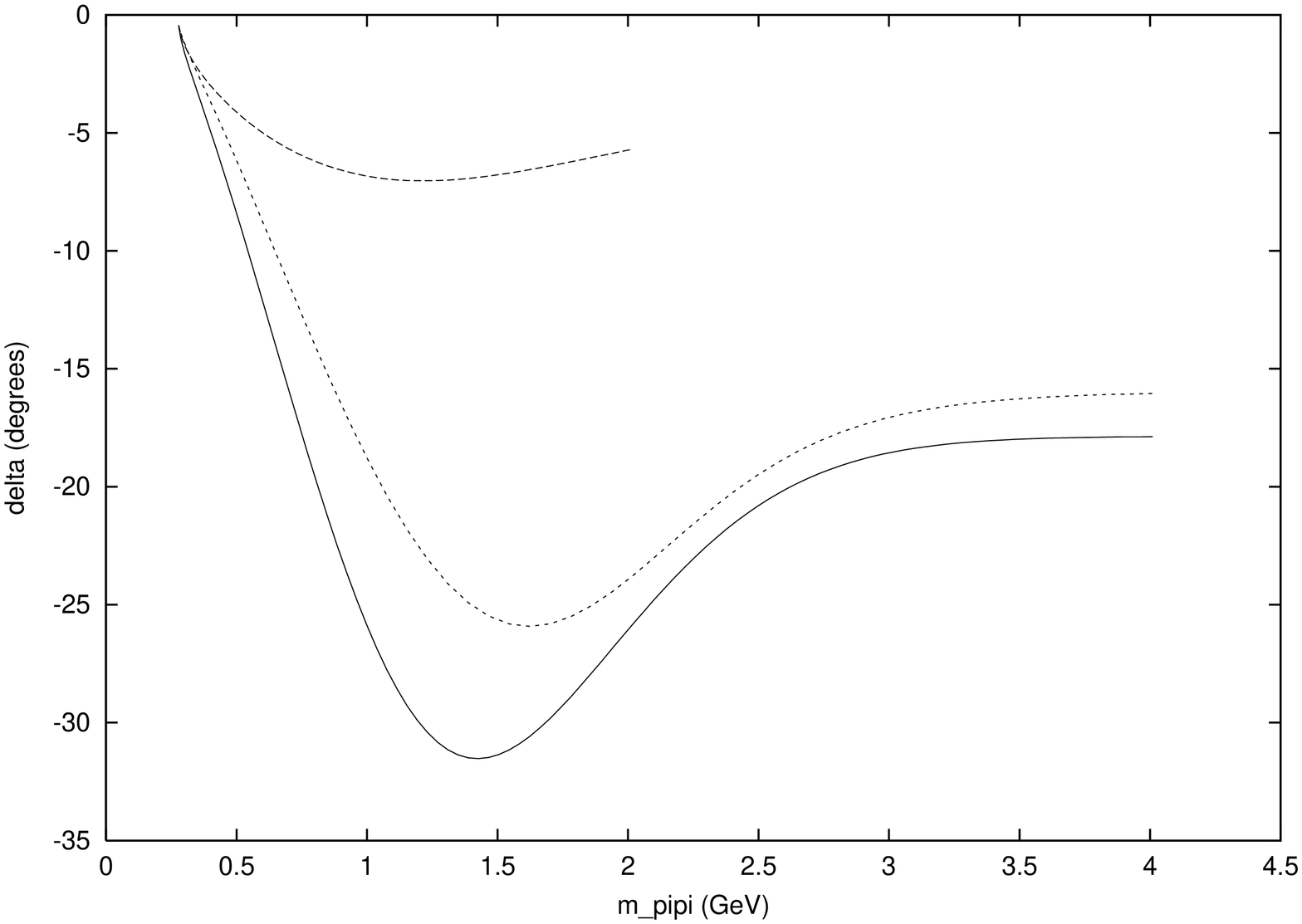}\\
   \parbox{14cm}
	{\centerline{\footnotesize 
	Fig.~4: I=2 $\pi\pi$ Scattering. The solid line is the QBD prediction, 
        the dashed line is the }}
	{\centerline{\footnotesize 
        nonrelativistic local prediction, while the
        dotted line is the nonlocal, relativistic prediction.}}
\end{center}

A related problem is the nonlocality of the $\pi\pi$ T-matrix 
(because composite particles are being scattered). Any extracted effective potential which
is local will induce unknown errors after being iterated. Fortunately, all three problems may
be dealt with by choosing to solve for the T-matrix directly in momentum space:

\begin{equation}
T_E(k',k) = V(k',k) + \int d^3p V(k',p) {1\over E - E(p) + i \epsilon} T_E(p,k)
\end{equation}

\noindent
As is evident from the equation, nonlocal effective potentials may be directly employed (in fact
these are proportional to the QBD T-matrices), unitarity is automatically restored, and 
nonrelativistic kinematics may be incorporated by use of the relativistic dispersion relation
for $E(p)$. This is illustrated as the dotted line in Fig 4, where one sees that the effect
of iterating the $\pi\pi$ potential is to weaken the scattering at higher invariant mass.

\subsection{Annihilation to Hybrids}

To this point attention has been restricted to `exotic' channels such as I=2, where resonance contributions
are forbidden. We now examine the issues involved in relaxing this constraint. 
Two possible intermediate states may be realized (at least at lowest order in the Fock space
expansion), annihilation to intermediate mesons or annihilation to intermediate hybrid mesons.
The latter process involves an intermediate state consisting
of a quark -- antiquark pair and a `gluon' (where the gluon may be an excited flux tube or a
constituent gluon, depending on one's picture of soft glue).

The nonrelativistic reduction of the one gluon exchange potential which
describes the coupling of
a $q\bar q$ pair to an intermediate perturbative gluon is\cite{Li:1994ys}

\begin{equation}
V_{ann} =   {2 \pi \alpha_s \over m^2} \left( {3\over 4} + \vec S_i \cdot \vec S_j\right)
\delta(\vec r_{ij}) {\lambda_i^a \over 2} {\lambda_j^a\over 2}.
\end{equation}

\noindent
However, the intermediate state in this case is a hybrid, which typically
has a mass some 1000 MeV above low lying mesons with the same quark content. 
Thus one may expect that it is more realistic to employ a gluon propagator
with a fictitious mass of roughly 1 GeV. One may incorporate this into
the expression above by multiplying it by a factor $f$ which is to be
fit to the data and which we expect to be roughly -1.
This rather speculative adjustment can be verified by 
comparing reactions with no annihilation to similar reactions where
annihilation is permitted. For example, I=2 $\pi\pi$ scattering may be
compared to I=0 $\pi\pi$; I=3/2 $K\pi$ scattering may be compared to
I=1/2 $K\pi$; and $K^+N$ scattering may be compared to $\pi N$ scattering. 
All indicate  that a negative value of $f$ is required; a fit yields
$f \approx -2.6$\cite{Li:1994ys}.

\subsection{Coupled Channels}

Even a cursory examination of typical hadronic scattering data reveals
the importance of intermediate resonance states. Unfortunately the
mechanism by which hadrons couple is poorly understood and surely 
involves complicated nonperturbative gluodynamics.
The current best guess is the purely phenomenological $^3P_0$ model
in which $q\bar q$ pairs are created with vacuum quantum numbers and
combine with the parent quarks to produce the daughter mesons.
Extensive calculations of meson and baryon decays have been made
with moderate success (typical errors in the amplitude are 20\% or less)\cite{3p0}.

Incorporating the $^3P_0$ model directly into the quark model is the
most direct way to include the effects of intermediate resonances.
This may be achieved most simply by writing the quark model in second
quantized notation and including a $^3P_0$  term which creates and annihilates
$q\bar q$ pairs. This is multiplied by a constant $\gamma$ which may 
be determined by comparison to a specific channel (say, $\rho \rightarrow \pi\pi$).

\begin{eqnarray}
\hat H &=& \int dx \,\big( -{\nabla^2 \over 2 m_q} b_x^\dagger b_x - {\nabla^2 \over 2 m_{\bar q}} d_x^\dagger d_x  \big) 
+ \gamma \int dx\,\big( b_x^\dagger \sigma \cdot {\buildrel \leftrightarrow \over \nabla}
d_x^\dagger +{\rm H.c.} \big).
\nonumber\\
&+& {1\over 2} \int dx \, dy \, \big( b_x^\dagger b_y^\dagger + d_x^\dagger d_y^\dagger\big) V(x-y) \big(b_y b_x+ d_y d_x\big).
\end{eqnarray}

\noindent
The field theory is simplified by restricting the Fock space to the
meson and meson-meson sectors of interest. Thus we
project onto $\vert A\rangle$, $\vert BC\rangle$ by making the following Ansatz for the
exact eigenstate:

\begin{eqnarray}
\vert \Psi \rangle &=& \int \varphi_A(r_1 - r_2) b_{1}^\dagger d_{2}^\dagger \vert 0\rangle \nonumber\\
&+&  \int \sum_{BC} \Psi_{BC}({r_2 + r_4 - r_1 - r_3\over 2}) \varphi_B(r_1-r_3) \varphi_C(r_2-r_4) b_1^\dagger d_3^\dagger b_2^\dagger d_4^\dagger \vert 0\rangle 
\end{eqnarray}

\noindent
Varying the reduced Hamiltonian matrix element with respect to the
unknown meson $\varphi_A$ and meson-meson $\Psi_{BC}$ wavefunctions yields
the coupled channel Schr\"odinger equation:


\begin{eqnarray}
E \varphi_A(r) &=& H_{q\bar q}(r) \varphi_A(r) \nonumber\\
-\gamma \int \vec \Sigma \cdot ( \nabla_B &+& \nabla_C + \nabla_{BC}) \varphi_{0B}(r/2-x) \varphi_{0C}(r/2+x) \Psi_{BC}(-r/2), \\
{-1\over 2 \mu_{13,24}} \nabla_R^2 &+& \int\int K_E(x,y,R) \Psi_{BC}(R') +
\int\int V_E(x,y,R) \Psi_{BC}(R') \nonumber\\
&-& 8 \gamma \int \vec\Sigma \cdot (\nabla_B + \nabla_C +\nabla_{BC}) \varphi_{0B} \varphi_{0C} \varphi_A(-2R) \nonumber \\
&=& E \Psi_{BC}(R) + E\int N_E(x,y,R)\Psi_{BC}(R')
\end{eqnarray}

\noindent
Here $r$ is the interquark radius in the meson channel and $R$ is the
intermeson distance in the meson-meson channel. Remarkably there is a simple
relationship between these coordinates: $R = r/2$.
$K_E$, $V_E$, and $N_E$ represent the exchange kinetic energy, potential, and
normalization kernels respectively. Wavefunctions with a `naught' subscript $\varphi_0$ 
represent mesonic wavefunctions without the effects of channel mixing (so that we have
assumed no wavefunction distortion in deriving this equation). The first of these equations is the
nonrelativistic quark model ($H_{q \bar q}$) supplemented with a term which
couples it to the meson-meson continuum -- thereby `unquenching' the quark model.
The second equation is the resonating group equation which describes meson-meson
scattering in the CQM (the Born order T-matrix for this equation is provided
by the QBD). The term proportional to $\gamma$ provides the desired coupling
to intermediate resonances.

Eqns (5,6) may be solved with standard coupled channel methods and the
effects of unquenching the quark model and of intermediate resonances in
scattering problems may be examined. An example of this is given in Fig 5
where the effect of coupling virtual $\rho$s  to the $\pi\pi$ P-wave channel
is studied.  The $\rho$ mass has been shifted down 80 MeV while the bare width is 10 MeV,
this  corresponds to an `RPA' width of 90 MeV\cite{Page:1999gz}.

\begin{center}
   \includegraphics[width=5cm,angle=-90]{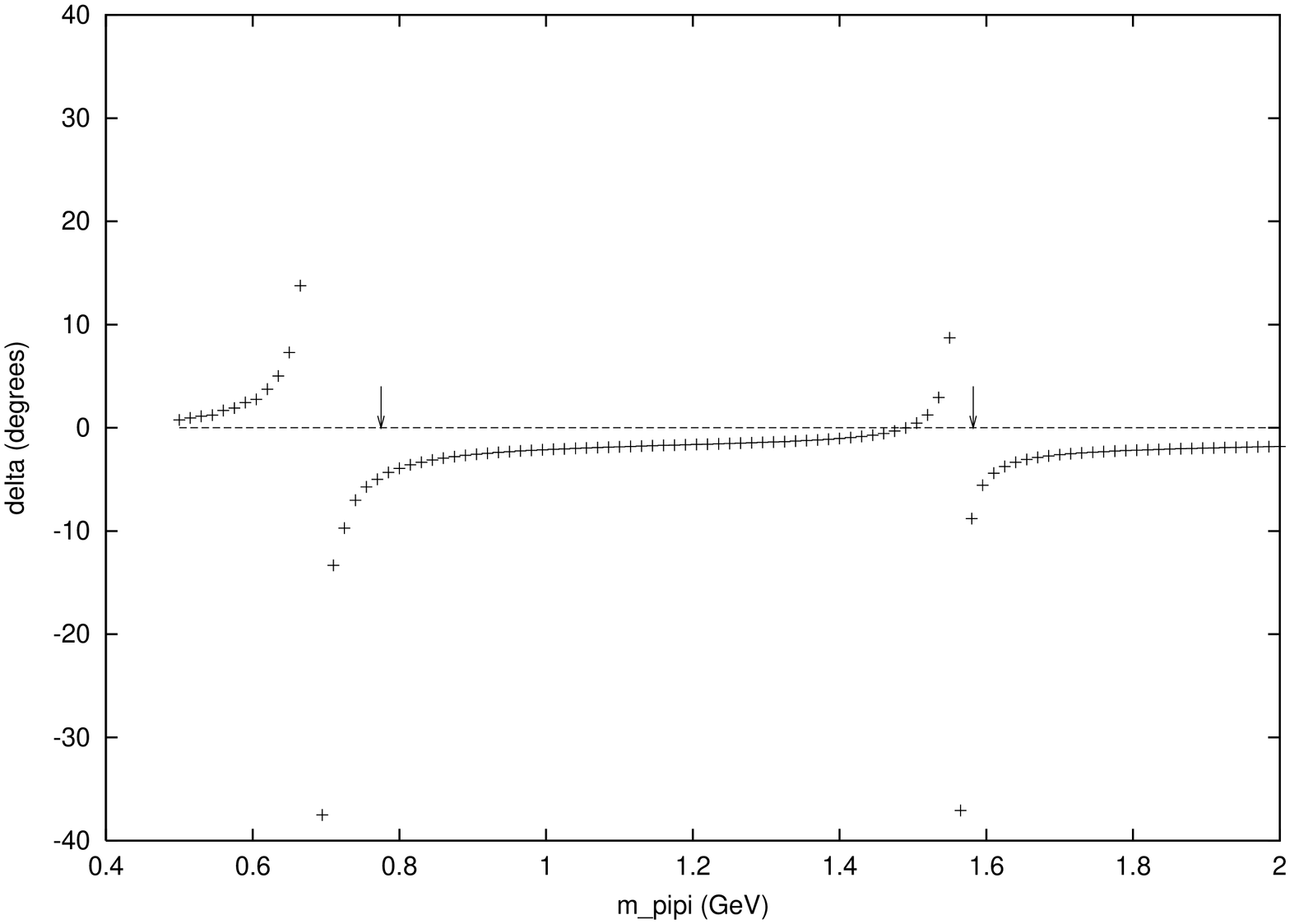}\\
   \parbox{14cm}
	{\centerline{\footnotesize 
        Fig.~5: I=1 $\pi\pi$ Scattering. The arrows indicate the locations of
        the `bare' $\rho$ and $\rho'$ mesons.}}
\end{center}

\subsection{Chiral Pions}

The apparently successful description of $\pi\pi$ scattering evident
in Fig 1 is perhaps surprising given the chiral nature of the pion
and its interactions at low energy. Explaining this success will go a long
way towards explaining the unwarranted success of the CQM in the light
quark sector\cite{Szczepaniak:2000bi}. One way to do this is to construct a
model of strong QCD which incorporates the physics of chiral symmetry
breaking at a microscopic level. In this way one may compute $\pi\pi$
scattering with composite particles while observing the dictates of
chiral symmetry. It is possible to construct such a model by assuming
a nontrivial QCD vacuum (typically a BCS-type vacuum) and building states
on this vacuum with the random phase approximation to the full Bethe Salpeter
equation. This has been done in Coulomb gauge QCD\cite{Szczepaniak:1997gb,Orsay},
with similar calculations in the Schwinger-Dyson approach \cite{Roberts:1994ks}

\section{Conclusions}

Studying and understanding hadronic interactions is vital to hadronic
physics and is an important part of nuclear and electroweak physics.
It is probable that a microscopic description of hadronic interactions
is necessary if one wishes to understand these phenomena in extreme
conditions or in poorly known channels. The constituent quark model
provides an excellent starting point for developing the understanding
required to construct a reliable model of strong QCD. In the meantime,
it also serves as an excellent phenomenological guide to the
interpretation of scattering experiments. 
The development of continuum field theoretic models capable of
describing hadrons and their interactions is in its
infancy -- we look forward to their maturation and application to
reaction processes.

The author thanks the organizers of the Hirschegg workshop for their labours and
for introducing the participants to the delights of the Kleinwalsertal. A large
portion of the work described here 
was performed with Ted Barnes and Adam Szczepaniak. The author is grateful
to Paul Geiger for discussions on coupled channels.
This work was supported by the US DOE under contracts
DE-FG02-00ER41135 and DE-AC05-84ER40150.


\begin{thebibliography}{99}
\bibitem{foot2}
This may be traced to the importance of the quark-level spin-orbit interaction.  Work
is underway to test this.
\end{thebibliography}
\end{document}